# The Analysis of Long-Term Frequency and Damping Wandering in Buildings Using the Random Decrement Technique

Ali Mikael - Philippe Gueguen -  Pierre-Yves Bard - Philippe Roux - Mickael Langlais

ISTerre - Université Joseph Fourier - Grenoble I - CNRS/IFSTTAR

Philippe Gueguen

ISTerre

Campus Universitaire

1381 rue de la piscine

38041 Grenoble Cedex 9

France

email: philippe.gueguen@ujf-grenoble.fr

**Abstract**

The characterization and monitoring of buildings is an issue that has attracted the interest of many sectors over the last two decades. With the increasing use of permanent, continuous and real-time networks, ambient vibrations can provide a simple tool for the identification of dynamic building parameters. This study is focused on the long-term variation of frequency and damping in several buildings, using the Random Decrement Technique (RDT). RDT provides a fast, robust and accurate long-term analysis and improves the reliability of frequency and damping measurements for structural health monitoring. This reveals particularly useful information in finding out precisely how far changes in modal parameters can be related to changes in physical properties. This paper highlights the reversible changes of the structure's dynamic parameters, correlated with external forces, such as temperature and exposure to the sun. Contrasting behaviors are observed, including correlation and anti-correlation with temperature variations.

**INTRODUCTION**

Over the last two decades, ambient vibrations methods for assessing the modal parameters of structures have received considerable attention. Since the design forces and displacements in structures are frequency and damping dependent (based on the seismic coefficient $C(T,\xi)$ where $T$ is the period of the building and $\xi$ is the damping ratio), the use of ambient vibrations methods provides relevant information on the elastic characteristics of a building at relatively low cost. Carder (1936) established the first empirical relationships between the height and period of buildings in the US, and recent papers (e.g., Farsi and Bard, 2004; Gallipolli et al., 2009; Michel et al., 2010) have provided other relationships based on experimental data for European buildings with different designs. In practice, semi-empirical formula are provided in earthquake-designed code or guidance documents, such as the ATC3-06 (ATC, 1978) in US or the EC8 in Europe (Eurocode 8, 2004), taking the form $T=C_tH^b$, T the period, H the height and $C_t$ and b depending on the building design and material. Frequency and damping are then key parameters for earthquake design and seismic vulnerability assessment since, as mentioned by Spence et al. (2003), the adjustment of structural models must assume a large set of unknown parameters influencing the response of existing buildings. Knowing frequency and damping can reduce the range of errors and epistemic uncertainties for representing the vulnerability as fragility curves (Michel et al., 2011).

Finally, ambient vibrations modal analysis methods provide an effective tool for short- and/or long-term monitoring of building health indicators, such as aging effects, or after extreme events. The basic idea is that any modification of the stiffness, mass or energy dissipation characteristics of a system may alter its dynamic response (Doebling et al., 1996; Farrar and Worden, 2007). Variations in these modal parameters can result from a change in the boundary conditions (e.g. fixed- or flexible-base structure), mechanical properties (e.g. reinforcement or retrofitting) or the elastic properties of the material (e.g. Young's modulus). The causes may also be related to non-linear responses of the buildings. For example, transient variations are observed during seismic excitation due to the non-linear response of the soil-structure boundary (e.g., Jennings and Kuroiwa, 1968; Crouse and Jennings, 1975; Todorovska and Trifunac, 2008, Todorovska, 2009) or to the closing/opening process of pre-existing cracks within the elements of the reinforced concrete buildings (e.g., Luco et al., 1987; Meli et al., 1998; Clinton et al., 2006; Michel and Gueguen, 2010). Permanent variations may also appear due to structural damage due to strong seismic motion (e.g., Safak and Celebi, 1991; Celebi et al., 1993; Dunand et al., 2006; Clinton et al., 2006). Finally, these variations can be long-term, reversible and slight, as recently observed by Clinton et al. (2006) and Todorovska and Al Rjoub (2006). In most cases, the fluctuations of the fundamental frequency of buildings was correlated to the temporal variations of the atmospheric conditions (temperature, humidity, etc.) influencing the properties of the building and the soil, i.e. modifying the soil-structure interaction and the boundary conditions. Similarly, most previous studies conducted in civil engineering structures (e.g., Clinton et al. 2006; Deraemaker et al., 2008; Hua et al., 2007; Nayeri et al., 2008) have shown that temperature is the most significant cause of variability of modal frequencies.

Farrar et al. (2010) mentioned that frequencies are probably the modal parameters most sensitive to change, particularly because the loss of stiffness directly impacts the frequency values. Nevertheless, the apparent damping coefficient recorded in the building may also be directly sensitive to a local loss of stiffness (e.g., Jeary, 1997; Modena et al., 1999) or to soil-structure interaction (e.g., Gueguen and Bard, 2005). Moreover, since the natural fluctuations of modal parameters has been reported by several authors, and since the variation in these parameters is used for structural health monitoring, two important points must be clarified: (1) what are the smallest variations in modal parameters that can be detected by ambient vibrations and related to structural health (in some cases, the amplitude of the variability of modal parameters caused by temperature may be such that it masks the changes due to physical damage); and (2) are variations in modal parameters always related to external conditions with the same trend regardless of the building typology?

The main purpose of this paper is to analyse the long-term variations in the frequency and damping coefficient of existing buildings using the Random Decrement Technique (RDT). We applied RDT to three existing buildings equipped with temporary or permanent instrumentation. After presenting the method in section Damping and frequency using random decrement technique, the variations in frequency and damping are discussed in section One-year observation of modal parameters, as well as the link between frequency and damping fluctuations and the effect of temperature. Concluding remarks are presented in section Discussion.

**DAMPING AND FREQUENCY USING THE RANDOM DECREMENT TECHNIQUE**

Salawu (1997) focuses on the fundamental frequency fluctuations to detect changes in buildings. This parameter is very sensitive and it can be easily tracked for structural health monitoring or for the long-term analysis of buildings. For instance, under the single-degree-of-freedom system (SDOF) assumption, the fundamental frequency is related to stiffness *K* and mass *M*, through the equation:

$$\omega = \sqrt{\frac{K}{M}} \tag{1}$$

where *ω=2π.f* is the resonant circular frequency of the building. To first order, we assume that the mass M remains constant. The frequency variations are then related to the variation of stiffness *K*, *K* depending on the properties of the building (e.g., Young's modulus, height, design of the building, etc.), but also on the cracks opening and the boundary conditions (e.g., fixed- or flexible-base building). Frequency analysis is currently included in building tests since the simple fast Fourier transform of ambient vibrations recorded at the top of the building provides relevant information on the modal frequencies, above all for very tall buildings exhibiting a good signal-to-noise ratio and a low damping. Many papers have discussed the fundamental frequency sensitivity in the case of damage after an extreme event (e.g. Luco et al., 1987; Celebi et al., 1996; Ivanovic et al., 2000;

Todorovska, 2009; Michel and Gueguen, 2010), on the long-term analysis of buildings (e.g., Clinton et al., 2006; Nayeri et al., 2008) or on the damage localisation (e.g., Kim and Stubbs, 2003; Xu et al., 2007), most studies being based on the Fourier analysis or the time-frequency analysis of the recorded data.

Even though the damping coefficient is a critical parameter for assessment design, it is not usually easy to determine the damping coefficient in experimental conditions and its physical origin in most practical systems are seldom fully understood. Jacobsen (1930) provided an early formulation for damping, corresponding for the sake of simplicity to an equivalent coefficient including viscous damping (proportional to velocity), constant Coulomb's damping and material damping. Crandall (1970) mentioned that damping is amplitude dependent, i.e. non-linear effects may be observed that increase the damping value as shaking amplitude increases. Moreover, and whatever the excitation amplitude, damping is frequency dependent and becomes significant at the highest resonant frequencies (Crandall, 1970). The shape of the frequency-response curve at each mode is controlled by the system's damping coefficient and can be estimated using the half-power bandwidth method (Clough and Penzien, 1993). The free-oscillating response of the SDOF, as controlled by the e$^{-\xi\omega t}$ function, is thus proportional to the frequency $\omega$ and the critical damping coefficient $\xi$. The critical damping coefficient $\xi$ is expressed as follows:

$$\xi = \frac{c}{2\sqrt{K M}} \qquad (2)$$

where $c$ represents the energy-loss mechanism, and $K$ and $M$ the stiffness and mass of the SDOF respectively. In buildings, radiative damping must be included in the "apparent" $c$, according to the soil-structure interaction since it affects the seismic energy radiated back into the ground (e.g., Gueguen and Bard, 2005; Gueguen et al., 2000, 2002). In very tall buildings, aerodynamic effects may also contribute to damping (Solari, 1996; Tamura and Suganuma, 1996). Jeary (1986), Lagomarsino (1993), Farsi and Bard (2004) and Satake et al. (2003) proposed empirical relationships for estimating damping as a function of height, frequency and shaking amplitude. These equations can be used to offer a rough estimate when designing new buildings. Several authors have analysed the variation of damping with shaking amplitude and frequency (Hart and Vasudevan, 1975; Jeary, 1986; Tamura and Suganuma,1996; Li et al., 2000; Satake et al., 2003). All observed that damping increases with shaking amplitude and increases with the fundamental frequency of the building (Satake et al., 2003). Finally, as previously mentioned for frequency, the variation of stiffness $K$ may also cause the damping coefficient to vary.

An effective solution to track frequency and damping variations over time is to apply the RDT. This method was first proposed by Cole in 1973 and is based on the fact that, at any given time, ambient vibrations contain a random and impulse element. By stacking a large number of windows with identical initial conditions, ambient vibrations remain stationary and the impulse response of the structure is revealed. Vandiver et al. (1982) and Asmussen et al. (1999) provide details on the theory of RDT and its mathematical formulation that can be simplified by:

$$RDT(\tau) = \frac{1}{N} \sum_{i=1}^{N} s(t_i + \tau)$$

(3)

where $N$ is the number of windows with fixed initial conditions, $s$ is the ambient vibration window of duration $\tau$, and $t_i$ is the time verifying the initial conditions (Fig. 1). The choice of initial conditions is a key point in ensuring the stability of the Random Decrement signature. The null displacement and positive velocity conditions proposed by Cole (1973) and verified by Asmussen et al. (1999) were used in the present work. The number of windows N is also critical to obtain a stable and relevant damping estimation. Jeary (1997) recommended at least 500 windows, i.e. one hour of recording, as sufficient for almost buildings having fundamental frequencies between 0.5 and 2Hz. In our study, the duration of each recording was one hour, corresponding to more than 1000 periods $T$ of the fundamental mode (around 1Hz for the buildings tested herein), keeping the same criteria whatever the building. We considered a sampling of one hour, i.e. one value of frequency and damping per hour. No overlap was considered between two successive one hour windows. $l$ is the length of the windows selected (and stacked) in one hour. In our case, since the RDT amplitude gets smaller towards the end of the signature, we limited its length to 20 seconds to avoid the average being biased by the reducing amplitude. The number of windows $N$ selected for one hour depends on the number of times the initial conditions are found on the one hour window. The minimal value is 3600/20, i.e. 180 windows without overlap. In practice, considerable overlap exists between each window of 20 seconds, the initial conditions $t_i$ ending up almost in every period, i.e. about 1 second. $N$ is then greater than 500 windows.

Before the RDT processing, a fourth-order band pass filter was applied to the raw data, centred on the expected fundamental frequency with a 10% frequency band. The RDT signature of the mode is exponentially damped and corresponds to the so-called logarithmic decrement of damping (Clough and Penzien, 1993). Its period is computed by averaging the time lapse between two upward zero crossing points. Simultaneously, we estimate the damping value $\xi$ from the logarithmic decrement of the RDT signal by adjusting an $e^{-\xi\omega t}$ function.

To demonstrate the RDT for the long-term monitoring of buildings, this method was first applied to the Mont-Blanc (MB) and Belledonne (BD) buildings (Fig. 2a), two of the three Ile Verte towers located in Grenoble (France). These stand-alone towers are 30-storey reinforced concrete buildings. They were built between 1963 and 1967 and were not designed to be earthquake resistant. Their shape is a rhombus of (LxT) 40x20m. The structural strength system is provided by two main continuous walls throughout the height of the structures (100m), completed by small RC walls in the horizontal directions. They have the same orientation, N10°. For sake of simplicity, their L and T directions are considered as North-South and East-West, respectively, throughout the paper. Soil conditions are soft, with a mixture of clay deposits and sand and gravel in thin layers, resulting from the glacio-lacustrine deposits and later fluviatile processes that are characteristic of Alpine valley deposits such as the Grenoble basin (Gueguen et al., 2007). Ambient vibrations were recorded simultaneously at the top of the two towers during one month using a 24 bit Analog/Digital CityShark$^{TM}$ acquisition system (Chatelain et al., 2000) connected to a 5-s Lennartz 3C velocimeter. Continuous recording was performed and recording files were divided into one-hour long time windows sampled at 50Hz. This sample rate was selected in order to optimize the acquisition for detecting the building frequencies (estimated using empirical relationships close to 1Hz) and to use the memory storage of the temporary experiment. The acquisition systems in the two buildings were completely independent.

Examples of fast Fourier transforms are shown in Fig. 2a (right). The behavior of the two buildings is quite similar, their fundamental mode being at 0.67Hz (T) and 0.89Hz (L), and 0.65Hz (T) and 0.84Hz (L) for the BD and MB buildings, respectively. Overtones of bending modes close to 3Hz and 5Hz were also observed, as well as a torsion mode close to 1Hz. All these values were also observed by Michel et al. (2011) using extensive modal analysis, with multi-channels recordings. Herein, only the fundamental mode was used to test the RDT. The long-term variations of frequency and damping were then computed hourly from the RDT, i.e. for time-windows greater than 1000 periods. Frequency fluctuations are shown in Fig. 3 for one month and for the two buildings and the two horizontal directions. Near-perfect synchronization was observed in the frequency variations between the two buildings (Fig. 3a), RDT being able to detect very small fluctuations (less than 0.1%). Stronger transient variations, such as between the 10th and 11th of August, were also observed for the two buildings. Since the buildings are completely independent, the origin of these variations must be physical and directly related to different building stiffness or boundary conditions. The trend roughly follows the regional temperature curves provided by a meteorological station 20km from the buildings. Correlations with the daily variations are clear and longer period of variations are also observed, such as during the second week of August. The same trend was previously observed by Clinton et al. (2006), i.e. frequency increases with temperature. The scientific explanation for frequency variations has not yet been completely understood but it may result from the expansion of concrete or cladding in relation to sun exposure. For example, Fig. 3b shows the differences between behavior in the NS (L) and EW (T) directions. Frequency variations are less correlated between T and L for either building (correlation coefficient $C_{BD}=0.75$ and $C_{MB}=0.65$), and more correlated between the buildings when the same direction is considered ($C_T=0.89$ and $C_L=0.85$). This can be explained by the effects of sun exposure, the same faces being exposed in the same manner for both structures. Unlike with Clinton et al. (2006) and Herak and Herak (2010), no clear effects of rain and wind were observed (Fig. 3a). In our case, their effects may not have been strong enough to influence frequency, even though the building is founded on soft soil. Todorovska and Al Rjoub (2006) explained the frequency variations of the Millikan Library during rainfall by modification of the soil-structure interaction. For the wind, the data are not enough for doing the link between wind and frequency variation, further analysis are needed for definitively conclude, especially since the temperature is the dominant (at the first order) factor controlling the wandering. Some regional effects may also introduce a bias between the temperature at the meteorological station and at the building. Nevertheless, strong correlation are observed with temperature, let us assume a moderate regional effect.

RDT is also used for long-term monitoring of the damping value (Fig. 4). No clear variations are observed in Fig. 4a, although the stability of the measurement may reflect the efficiency of RDT for damping estimation. One conclusion could be that damping is less sensitive to external conditions. In contrast to results obtained with stronger motion (e.g., Hart and Vasudevan, 1975; Jeary, 1986; Tamura and Suganuma,1996; Li et al., 2000; Satake et al., 2003), the frequency and damping are not clearly correlated at these levels of excitation (Fig. 4b), regardless of the building and the direction. Finally, the relationship between the modal parameters (frequency and damping) and the temperature is displayed in Fig. 4c. We observe an overall positive trend between frequency and temperature. The highest correlation parameter was observed for the T direction ($C_{BD/T}$=0.56 - $C_{MB/T}$=0.48 - $C_{BD/L}$=0.36 - $C_{MB/L}$=0.09), i.e. the direction of the longest facade, confirming the effects of exposure to sunlight. No clear correlation was observed for damping.

In conclusion, one clear observation is the accuracy of RDT that enables the monitoring of very small variations (~0.1 %) in modal frequencies. RDT provides the effective assessment of the building's frequency and damping, the accuracy of the measurement having coefficient of variation ($\sigma/\mu_f$) close to 0.3% (Tab 1).
These variations are of physical origin, related to the natural variations in building stiffness. In these examples, temperature is the main parameter controlling the fluctuations. Using RDT, structural health monitoring can be achieved by distinguishing the natural (weather-related) frequency fluctuations from variations caused by natural aging or damage. Greater scattering are observed for damping than for frequency, but remain less than 10%, with $\sigma/\mu_\xi$ between 9 and 14%, assuming effective and stable assessment of the damping value by RDT (Tab. 1).

**ONE-YEAR OBSERVATION OF MODAL PARAMETERS**

In order to confirm the relevancy of RDT for long-term monitoring, two other buildings were tested (Fig. 2b): Grenoble city hall (CHB) and the Ophite tower (OT) in Lourdes (southwestern France). These two RC buildings are permanently monitored by accelerometric stations as part of the National Building Array program of the French Accelerometric Network (RAP, Pequegnat et al., 2008). Accelerometric sensors (EST-FBA) at the top of the building perform continuous recordings at 125Hz and transmit them in real time to the RAP National Data Centre hosted at the Institute of Earth Science (ISTerre, Grenoble). These structures were not designed to resist earthquakes and the original design report was not available to the authors.

The city hall building (CHB), fully described in Michel et al. (2010, 2011), is a 13-storey RC building (Fig. 2b), built in 1967 (LxTxH=44x13x52m). The inter-storey height is regular between the 3rd and 12th floors (3.2m) and larger on the 1st (4.68m) and 2nd stories (8m), above which there is a pre-stressed slab with a 23m span, supported by two inner cores. These cores, consisting of RC shear walls, enclose the stairwells and lift shafts at the two opposite sides of the building and provide the lateral resistance system (Michel et al. 2010). This building is founded on soft sediment, composed of mud and soft clay; its foundations are deep piles anchored on a stiff sandy layer at a depth of around -15m. The orientation of the building is close to N°45, i.e. the two faces of the building are exposed to the sun for the same period every day. Michel et al. (2010) performed a modal analysis of the building: the fundamental modes in the L and T directions are 1.16Hz and 1.22Hz respectively, with a torsion mode at 1.45Hz. Overtones were also observed, confirmed by a 3D Finite-Element model of the building (Michel et al., 2010), but they are not energetic enough to be analyzed here. Glass cladding is present, attached to a steel frame on the external perimeter of the structure.

The Ophite tower (OT) is an 18-storey dwelling building built in 1972 (Fig. 2c). The inter-storey height is regular (3.3m) throughout the height, with a regular lateral resistance system in the two horizontal directions, provided by RC shear walls. Its dimensions are 19m width (T), and 24m long (L). This building is founded on a rocky site, composed by Ophite rock, and we therefore assume a shallow foundation system. In addition, a temperature sensor is installed at the top of the OT which provides hourly temperature measurements. These values are influenced by the heat radiated by the building, thus reducing the amplitude of the temperature variations over the year. The orientation of the building is N°15, i.e. the L direction (close to the North-South direction) is more exposed to the sun. The fundamental frequencies of the building are 1.74Hz and 1.73Hz for the L and T directions respectively. Second bending modes were also observed, corresponding to 5.28Hz and 6.13Hz in the L and T directions respectively. An additional torsion mode was observed close to 2.2Hz, but this is not considered in this paper.

Two different responses were observed for the two buildings (Fig. 5). Frequency wanderings are clearly related to temperature, variations (see Tab. 1) being less significant for the CHB building ($\sigma/\mu_f$ close to 0.4%) than for the OT building ($\sigma/\mu_f$ close to 1.0%). The behavior of the CHB building seems to be the opposite of that observed in the BD and MB buildings, i.e. the frequency decreases when the temperature increases. This is also clear during the winter season (end of the time series) during which strong and fast temperature variations were observed at the OT and CHB building. During the winter, the frequency increases with the decrease of the temperature for the CHB and with the increase of the temperature for the OT for several clearly observed cycles of temperature variations. Moreover, a sudden response of the structure behavior was observed (as during december 2009 for the CHB) and we assume that the cladding effects must dominate in this case because of a slower diffusive effect of the temperature into the building. While for the CHB building, the L and T directions have similar variation amplitudes, the L direction of the OT building is more sensitive to heat than the T direction, which can be explained by the orientation, i.e. N°45 and N°15 for the CHB and OT buildings respectively. Moreover, the different trends observed between both buildings could be explained by the cladding effects and/or by the soil-structure interaction, since the two buildings have different cladding and different foundations, in spite of having the same resistance design (RC shear walls).

For damping, no clear correlation was observed for the CHB, with mean values around 1.027% ($\sigma/\mu_\xi$=11.88%) and 1.034% ($\sigma/\mu_\xi$=13.93%) in the L and T directions respectively (Tab. 1). This was not the case for the OT building, for which the trend in damping variations was opposite to those of frequency and temperature, as also observed by Clinton et al. (2006) and Herak and Herak (2010). The mean values of damping were 0.880% ($\sigma/\mu_\xi$=24.55%) and 0.827% ($\sigma/\mu_\xi$=15.96%) in the L and T directions respectively (Tab. 1).

This observation is also confirmed by Fig. 6 where frequency and damping are displayed for the CHB (Fig. 6a) and OT (Fig. 6b) buildings. For the latter building, the analysis of the second modes in the T and L directions are also shown, since the frequency modes were clearly detected using ambient vibrations. For the CHB, we observe a decrease in frequency as temperature increased ($C_L$= -0.6 - $C_T$= -0.6), the two faces of the building having the same exposure effect. This trend is opposite to that observed by Clinton et al. (2006) and Herak and Herak (2010), and for the two BD and MB twin towers too. No correlations are clearly put in evidence for the damping value ($C_L$=0.02 - $C_T$=0.12). On the contrary, the OT building showed a clear correlation between frequency and temperature, and also between damping and temperature, for the first and second modes (Fig. 6b). A change in trend was observed between the coldest (November to April) and warmest (May to October) months: the correlation parameter was $C_{L1}$= 0.69 and $C_{T1}$=0.84 (coldest) and $C_{L1}$= -0.32 and $C_{T1}$=0.51 (warmest). The same difference but opposite direction was observed for damping between the coldest and warmest months, i.e. damping decreased when temperature increased: $C_{L1}$= -0.74 and $C_{T1}$=-0.56 for the coldest months, and $C_{L1}$=

0.04 and $C_{T1}=-0.06$ for the warmest months. Frequency and damping must therefore be strongly correlated. The cladding effect could explain the first trend, i.e. an increase in stiffness related to the external windows exposed to the sun, also confirmed by the building orientation (N°15), while the second trend could be directly related to the effects of the concrete being exposed to the sun, reducing its Young's modulus. Similar trends were observed for the second mode of the OT building (Fig. 6b): for frequency, correlation coefficients were $C_{2L}=0.78$ and $C_{2T}=0.83$ for the coldest months and $C_{2L}=-0.07$ and $C_{2T}=0.62$ for the warmest; for damping, $C_{2L}=-0.5$ and $C_{2T}=-0.23$ for the coldest months and $C_{2L}=-0.09$ and $C_{2T}=0.43$ for the warmest.

All these observations are consistent with the theory that the variations are physical in origin, apparently related to the temperature's effects on the cladding. Moreover, the Fourier transforms of temperature, frequency and damping are displayed in Fig. 7. A clear peak is observed for the CHB and OT buildings, corresponding to a period of about 24 hours, confirming a direct correlation with the daily variation of temperature.

**DISCUSSION**

The natural fluctuations of the two main parameters that classically control the seismic response of a building during an earthquake (i.e., frequency and damping) have been analysed under ambient noise solicitation. While frequency variations had already been studied (Clinton et al., 2006; Hezak 2010), few works had focused on damping. The use of RDT has shown its effectiveness in measuring damping and frequency variations for the long-term monitoring of buildings. In our study, the main parameter controlling the fluctuations is the outside temperature. This was confirmed by the analysis of the Ile Verte towers, two stand-alone buildings (BD and MB) with the same design, that showed the same frequency versus temperature trend, based on independent measurements and for independent buildings. Conclusion is that these variations are of physical origin. By comparing horizontal direction behavior with respect to the building orientation, exposure to the sun seems to have a direct effect, influencing the overall stiffness of the building by acting on the cladding or external windows. No direct correlation with rain and wind was observed here due to a lack of strong atmospheric events.

While almost all previous studies showed a positive correlation between frequency and temperature (i.e., Clinton et al., 2006, Herak and Herak, 2010), no definitive conclusions can be drawn from this analysis. As a matter of fact, the frequency-vs-temperature trend seems to depend on the building. The responses of the MB and BD buildings exhibit a positive correlation with temperature, CHB an anti-correlation and finally the OT building showed two different mechanisms, with a trend that changed between the coldest and warmest months. Additional data will be needed to draw definitive conclusions on the physical parameters explaining these trends. We can suggest some explanations. All the buildings have different foundations, OT being on a rocky site while the MB, BD and CHB buildings rest on soft sediments. The soil-structure interaction affects the sensitivity of the building's response to external conditions. Todorovska and Al Rjoub (2006) have put forward such an explanation in the case of exceptional rainfall influencing soil properties. Cladding also differs from one building to another and the slight variations in frequency and damping could be directly related to this difference.

Nevertheless, RDT is an effective and robust tool that enables an accurate monitoring of the building response to ambient noise. The small variations observed are not related to the integrity of the building, assuming that less than 0.5% variation of the building fundamental frequency cannot be related to structural health but only to the natural fluctuations. This information is thus relevant for building monitoring, especially after extreme events when building tests using ambient vibrations are performed and compared with pre-event characteristics. Moreover, the modal parameters estimation using ambient vibration methods may be useful for reducing the model's uncertainties which are included in the fragility curves for seismic vulnerability (e.g. Michel et al., 2011). Figure 8 shows the normal and lognormal distribution of frequency and damping for the CHB and OT buildings. In both cases, σ was less than 0.01Hz and 0.2% for the frequency and damping values respectively. Nayeri et al. (2008) studied the variations of modal frequencies and damping in the Factor Building, a 17-storey steel frame structure located on the UCLA campus in California. They used a sophisticated method for modal analysis applied to ambient vibrations recordings over 50 days. The order of magnitude of the coefficient of variation σ/µ computed for the first mode was 1% with σ = 0.005, and close to 60% for damping with σ = 2.0. In the present work, RDT provided lower frequency variations (less than 0.3%) with σ =0.02 and of the damping coefficient (less than 15%) with σ = 0.2. These lower variations may be due to the type of building (steel frame for the Factor Building and RC shear walls for the buildings in our study) but may also be due to a more accurate assessment using RDT. Thanks to the fast operability of the RDT approach, long-term analysis can be envisaged for the permanent monitoring of buildings. Moreover, the effect of the frequency and damping variation must be introducing into models for seismic vulnerability assessment. The RDT provides an accurate estimate of these parameters, reducing then the epistemic uncertainties to the model (Spence et al., 2003). This method opens new perspectives for engineering seismology, well adapted to continuous and real-time data that start to be more and more available.

**DATA AND RESOURCES**

Accelerometric data used in this study were collected as part of the National Data Center of the French Accelerometric Network (RAP-NDC). Data can be obtained from the RAP-NDC at http://www‿rap.obs.ujf‿grenoble.fr/ (last accessed January 2011), part of the French Seismological and Geodetic Network (RESIF, http://www.resif.fr).

**ACKNOWLEDGMENTS**

This work was supported by the French Research National Agency (ANR) under the RiskNat program (project URBASIS n°ANR-09-RISK-009). Data were prepared with the help of Catherine Pequegnat, in the framework of the RESIF datacenter (http://www.resif.fr).

Table 1. Mean value, standard deviation and coefficient of variation of the frequency and damping obtained in the buildings using the Random Decrement Technique.

| Bldg. | L direction | | | | | | T direction | | | | | |
|---|---|---|---|---|---|---|---|---|---|---|---|---|
| | Frequency Hz | | | Damping % | | | Frequency Hz | | | Damping % | | |
| | $\mu_f$ | $\sigma$ | $\sigma/\mu_f$ | $\mu_\xi$ | $\sigma$ | $\sigma/\mu_\xi$ | $\mu_f$ | $\sigma$ | $\sigma/\mu_f$ | $\mu_\xi$ | $\sigma$ | $\sigma/\mu_\xi$ |
| BD | 0.890 | 0.003 | 0.33 | 1.094 | 0.099 | 9.05 | 0.672 | 0.002 | 0.29 | 0.666 | 0.073 | 10.96 |
| MB | 0.837 | 0.003 | 0.36 | 0.855 | 0.099 | 11.57 | 0.648 | 0.002 | 0.31 | 0.586 | 0.081 | 13.82 |
| CHB | 1.162 | 0.005 | 0.43 | 1.027 | 0.122 | 11.88 | 1.224 | 0.004 | 0.33 | 1.034 | 0.144 | 13.93 |
| OT | 1.747 | 0.013 | 0.74 | 0.880 | 0.216 | 24.55 | 1.741 | 0.022 | 1.26 | 0.827 | 0.132 | 15.96 |

*Note: μ: mean value; σ: standard deviation; σ/μ: coefficient of variation; BD: Belledonne building; MB: Mont-Blanc building; CHB: City-Hall of Grenoble; OT: Ophite tower; index f: frequency; index ξ: damping; L: longitudinal direction; T: transverse direction.*

Figure 1: Schematic view of the Random Decrement Technique (RDT) used in this paper. The blue circle correspond to the time respecting the initial conditions.

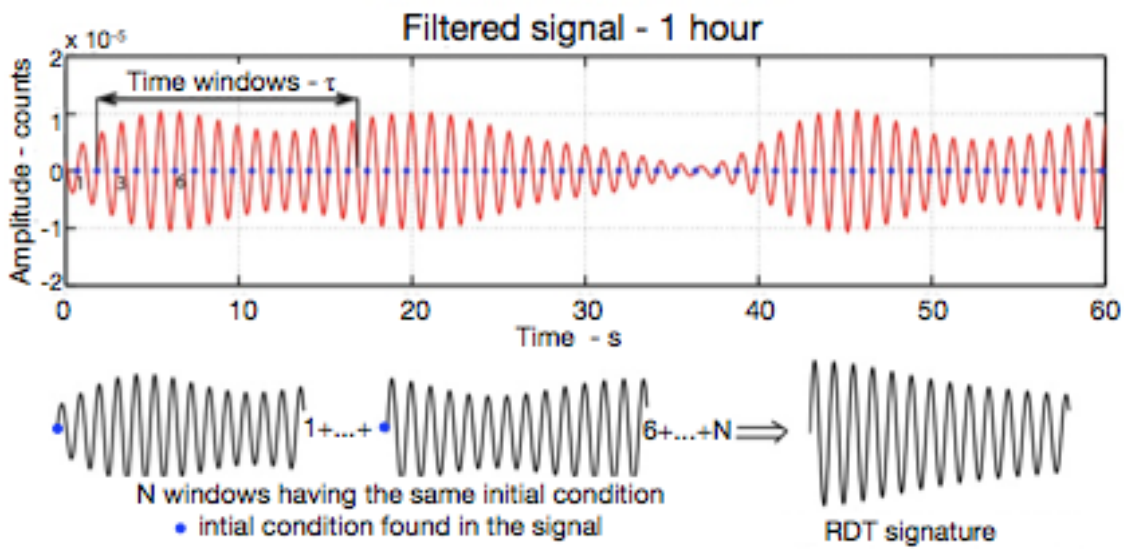

Figure 2: Spectral analysis of the buildings of this study, using ambient vibrations recorded at the top. a: Ile Verte Towers (Mont-Blanc MB and Belledonne BD). b: City Hall building of Grenoble (CHB). c: Ophite tower in Lourdes (OT). Spectra are averaged over 30 minutes.

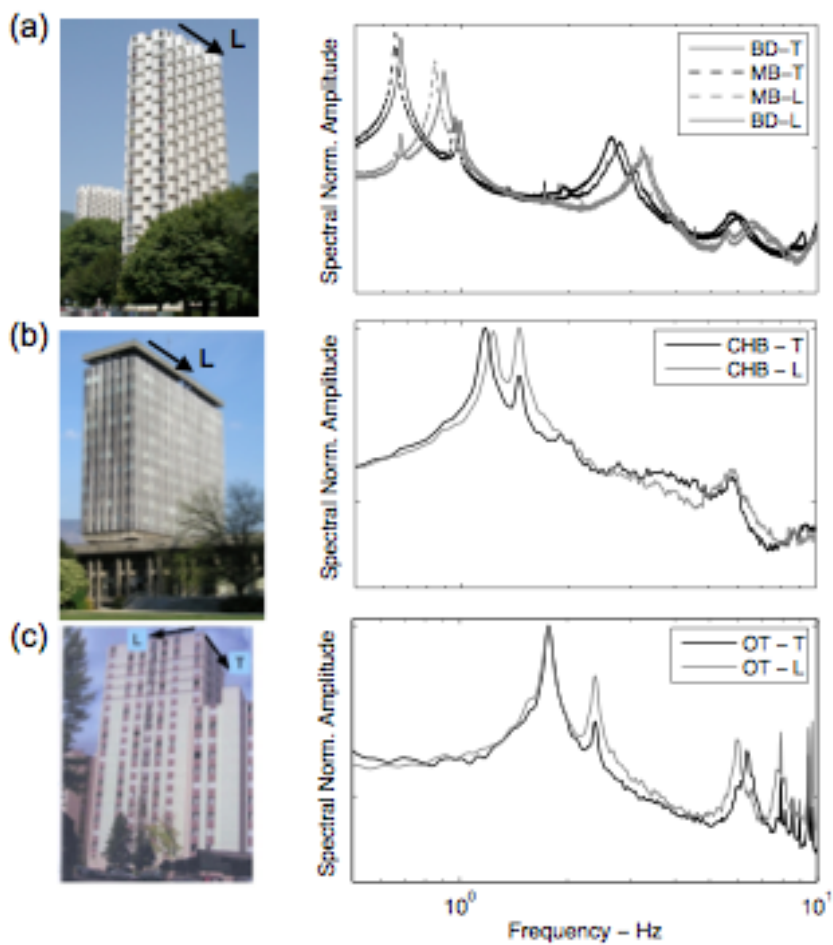

Figure 3: **a.** Frequency variations (normalized) of the Mont-Blanc (MB) and Belledonne (BD) buildings in the longitudinal L (upper row) and transverse T (middle row) directions for one month obtained using the Random Decrement Technique (the red line corresponds to temperature variations). The lower row corresponds to the rain and wind variations provided by a meteorological station, 20km from the buildings. **b.** Correlation between L and T directions of each building (left) and between BD and MB buildings for each component.

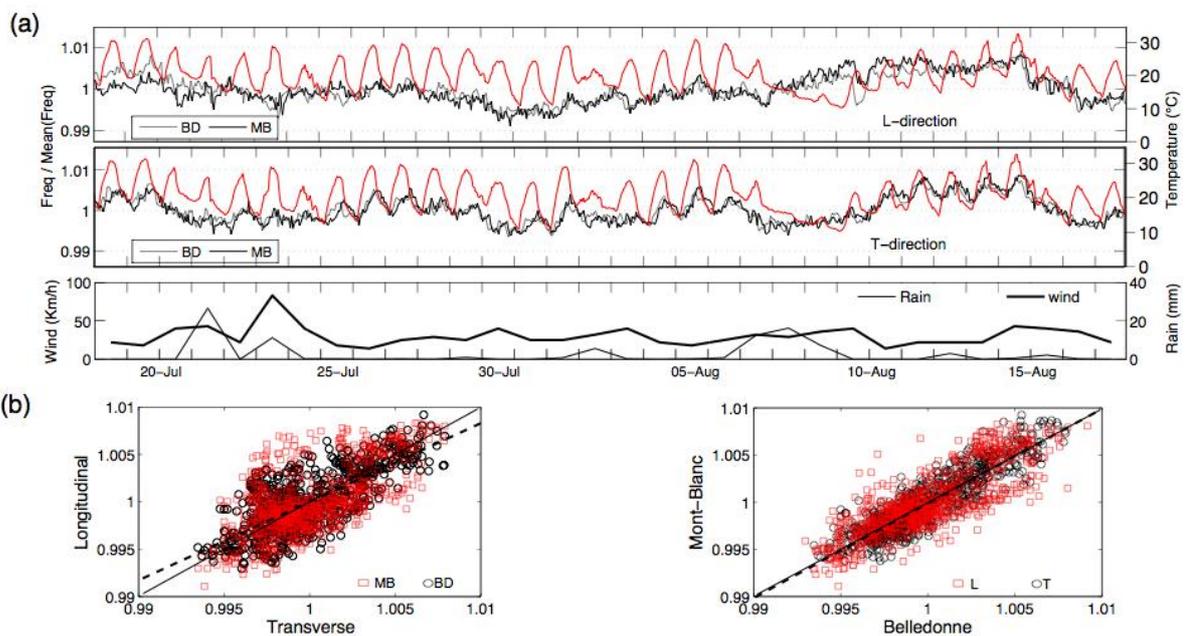

Figure 4: **a.** Damping variations of the Mont-Blanc (MB) and Belledonne (BD) buildings in the longitudinal L (upper row) and transverse T (middle row) directions for one month obtained using the Random Decrement Technique. **b.** Correlation between normalized frequency and damping for the MB (left) and BD (right) buildings. **c.** Correlation between temperature and frequency (left) and damping (right) variations for both buildings and directions. Mean values are removed for variation analysis.

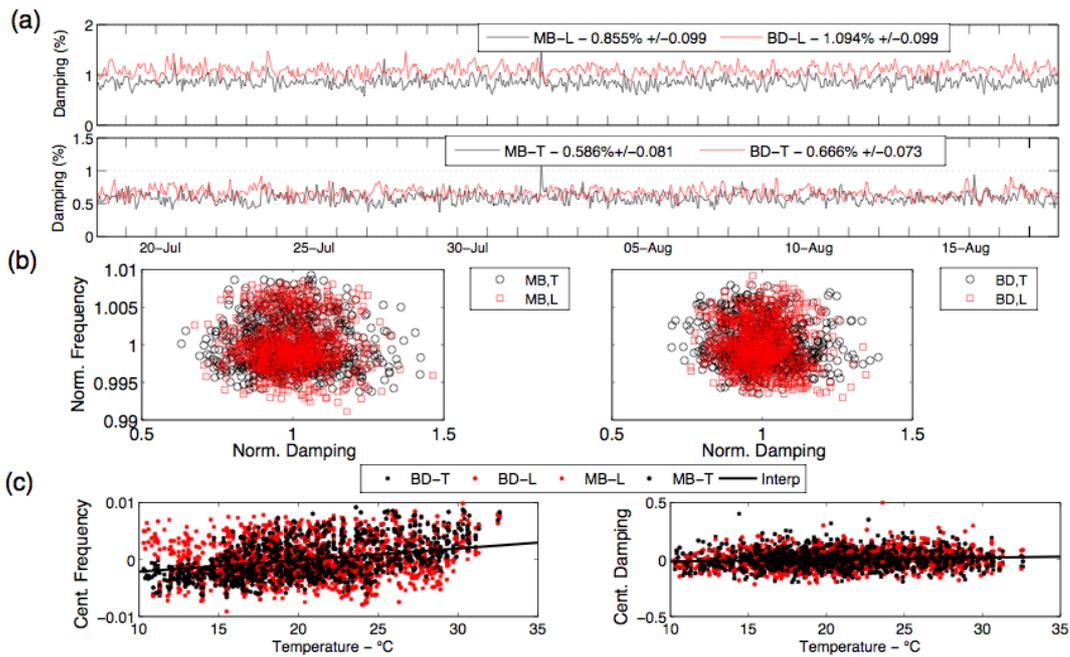

Figure 5: Frequency (**a**) and damping (**b**) variations for the City-Hall CHB (left) and Ophite OT (right) buildings over one year. The red line corresponds to temperature variations. The upper and lower rows correspond to the longitudinal (L) and transverse (T) directions, respectively. Continuous dotted lines indicate the lack of data.

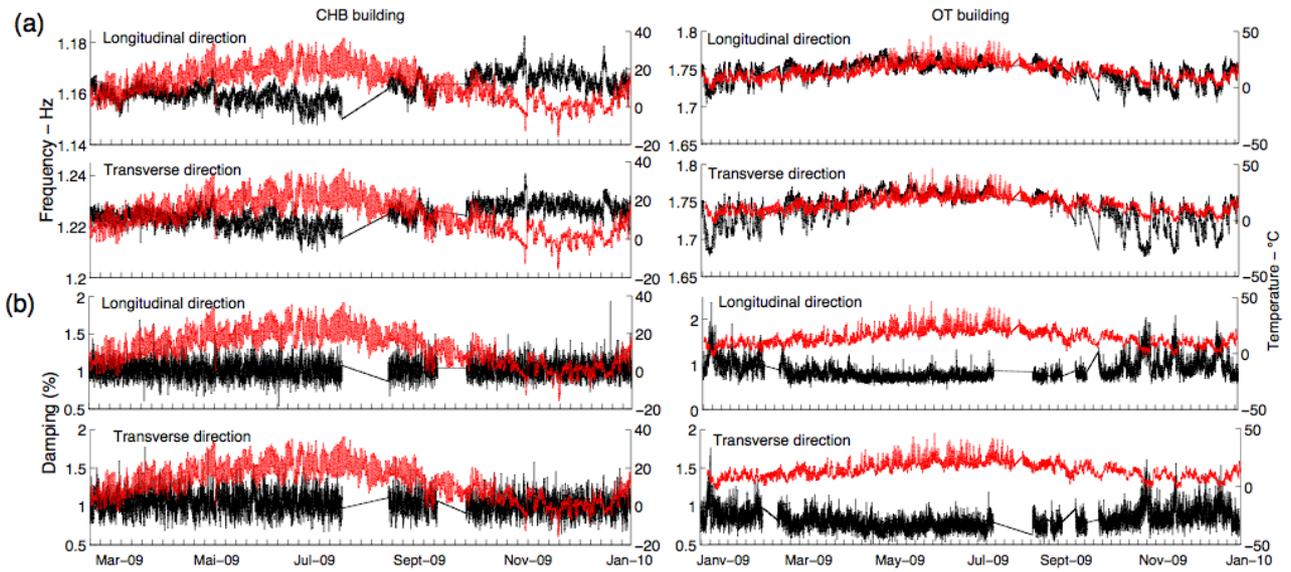

Figure 6: Correlation between frequency and damping variations with temperature in the longitudinal L (left column) and transverse T (right column) directions, for the City-Hall CHB (a) and Ophite OT (b) buildings. For the OT building, the first and the second modes are displayed.

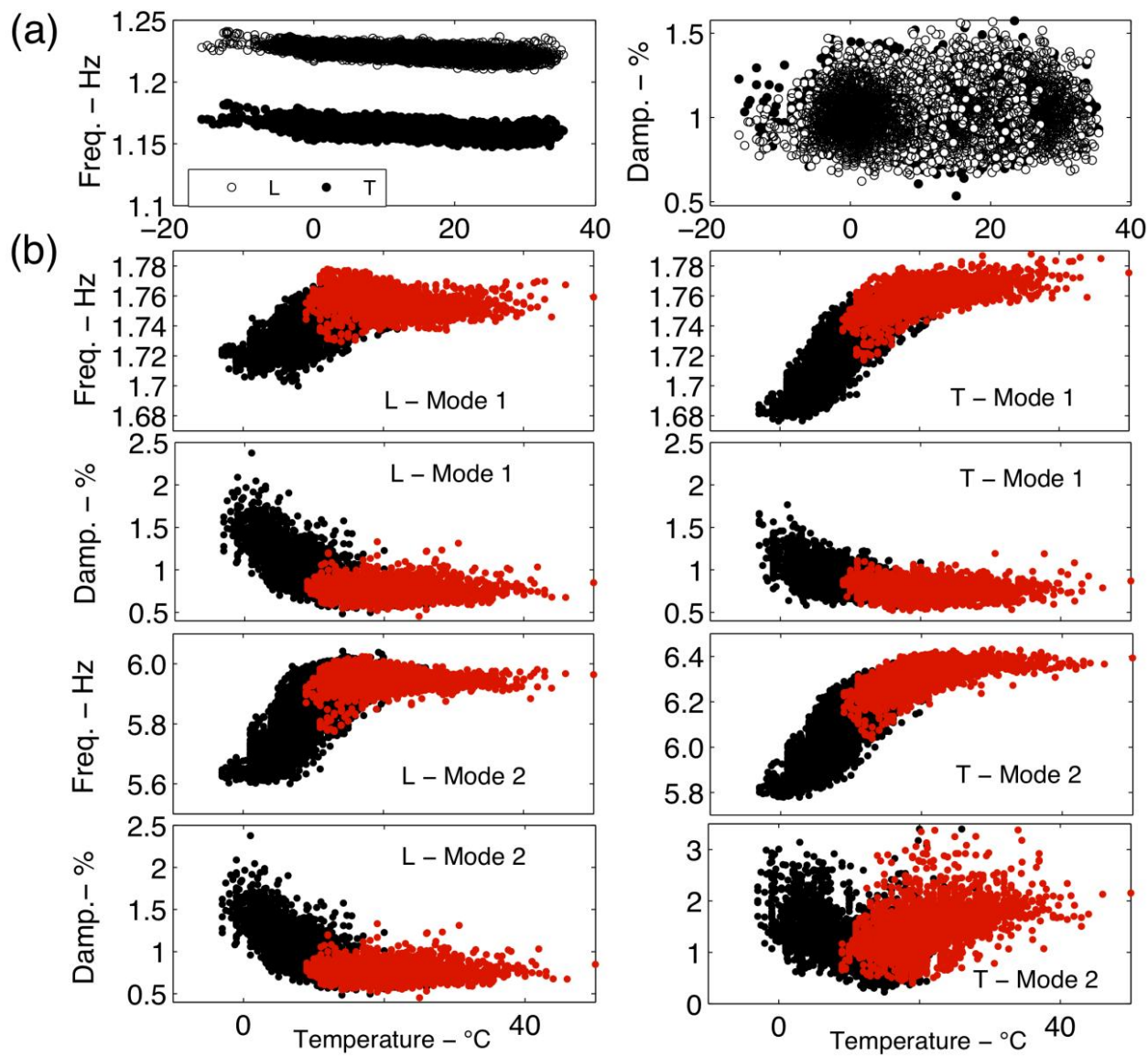

Figure 7: Fourier transform of the one-year temperature (upper row), building frequency (middle row) and damping (lower row) variations for the City-Hall CHB (left) and Ophite OT (right) buildings. Fourier spectra are computed by averaging 1024 points time windows (sampling rate: 1 point/hour)

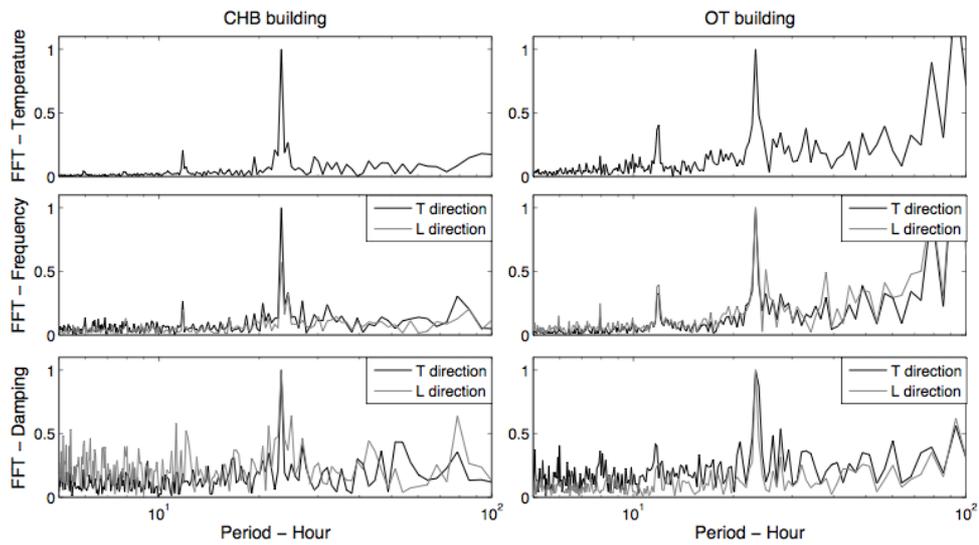

Figure 8: Lognormal (L, black line) and normal (N, dashed grey line) distribution adjusted to the frequency and damping values of the City-Hall CHB (upper part) and Ophite OT (lower part) buildings in the transverse T (left) and longitudinal L (right) directions.

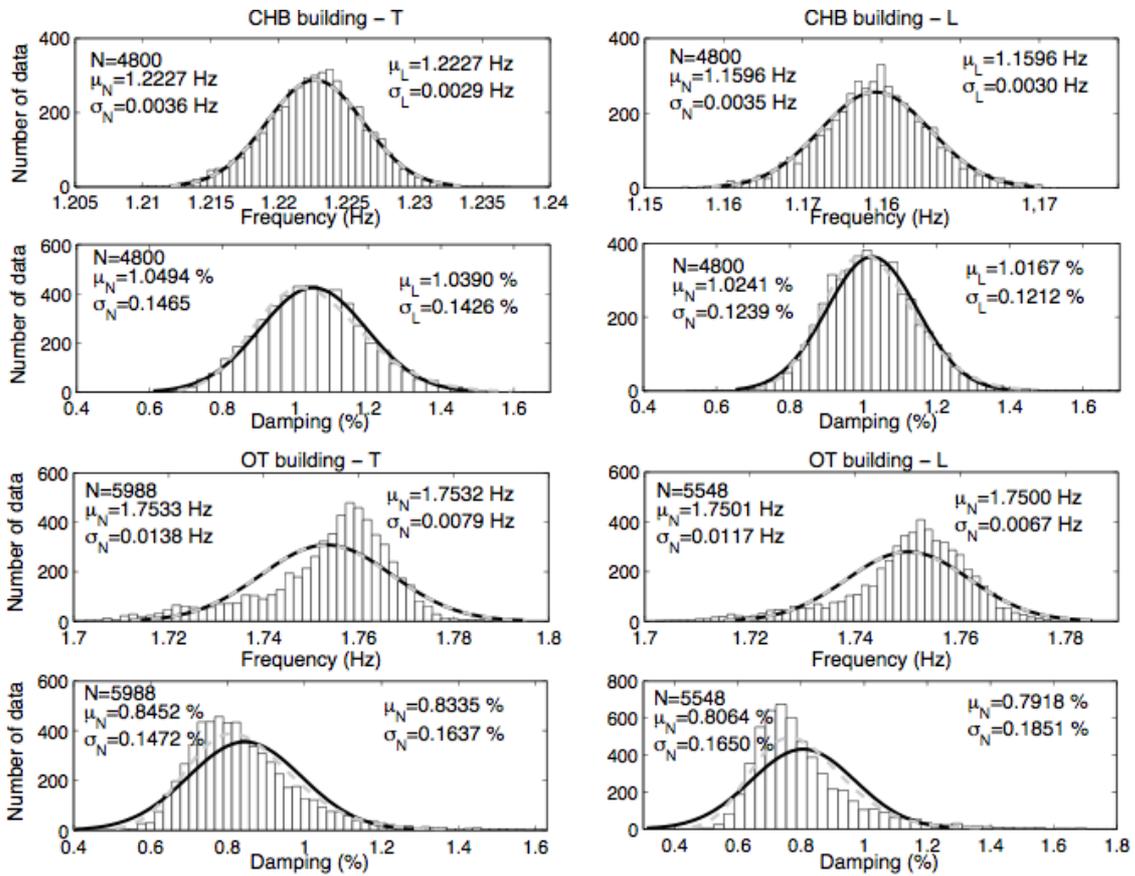